\documentclass[a4paper,11pt]{article}

\pdfoutput=1

\usepackage{jcappub_ver}
\usepackage{amsmath}
\usepackage{graphicx}

\usepackage[english]{babel}
\usepackage[utf8]{inputenc}
\usepackage{pdflscape}
\usepackage{enumerate}
\usepackage{amsbsy}
\usepackage{amsmath} 
\usepackage{graphics}
\usepackage{mathrsfs}
\usepackage{wrapfig}
\usepackage{mathtools}

\usepackage{amsfonts}
\usepackage{pstricks}
\usepackage{color}
\usepackage{setspace}
\usepackage{tensor}

\newcommand{\bea}{\begin{eqnarray}} \newcommand{\eea}{\end{eqnarray}}
\newcommand{\el}{\nonumber \\}
\newcommand{\re}[1]{(\ref{#1})}

\newcommand{\pat}{\partial}

\renewcommand{\sec}[1]{section \ref{#1}}
\newcommand{\fig}[1]{figure \ref{#1}}

\newcommand{\para}{\paragraph}

\renewcommand{\a}{\alpha}
\renewcommand{\b}{\beta}
\renewcommand{\c}{\gamma}
\renewcommand{\d}{\delta}

\renewcommand{\l}{\lambda}

\newcommand{\ha}{\frac{1}{2}}

\newcommand{\rmd}{\mathrm{d}}
\newcommand{\diag}{\mathrm{diag}}

\newcommand{\ie}{i.e.\ }
\newcommand{\eg}{e.g.\ }

\title{Higgs inflation with the Holst \\ and the Nieh--Yan term}

\author[a]{Miklos L\aa ngvik,}
\author[b]{Juha-Matti Ojanper\"{a},}
\author[b]{Sami Raatikainen}
\author[b,c]{and Syksy R\"{a}s\"{a}nen}

\affiliation[a]{\AA sh\"ojdens grundskola, \\ Sturegatan 6, 00510 Helsingfors, Finland}
	
\affiliation[b]{University of Helsinki, Department of Physics and Helsinki Institute of Physics,\\ P.O. Box 64, FIN-00014 University of Helsinki, Finland}

\affiliation[c]{Birzeit University, Department of Physics \\
P.O. Box 14, Birzeit, West Bank, Palestine}

\emailAdd{miklos.langvik@protonmail.com}
\emailAdd{juha-matti.ojanpera@helsinki.fi}
\emailAdd{sami.raatikainen@helsinki.fi}
\emailAdd{syksy.rasanen@iki.fi}

\abstract{
The action of loop quantum gravity includes the Holst term and/or the Nieh--Yan term in addition to the Ricci scalar. These terms are expected to couple non-minimally to the Higgs. Thus the Holst and Nieh--Yan terms contribute to the classical equations of motion, and they can have a significant impact on inflation.

We derive inflationary predictions in the parameter space of the non-minimal couplings, including non-minimally coupled terms up to dimension 4. Successful inflation is possible even with zero or negative coupling of the Ricci scalar. Notably, inflation supported by the non-minimally coupled Holst term alone gives almost the same observables as the original metric formulation plateau Higgs inflation. A non-minimally coupled Nieh--Yan term alone cannot give successful inflation. When all three terms are considered, the predictions for the spectral index and tensor-to-scalar ratio span almost the whole range probed by upcoming experiments. This is not true for the running of the spectral index, and many cases are highly tuned.
}

\begin{document}

\begin{flushleft}
	\hfill		 HIP-2020-22/TH \\
\end{flushleft}

\maketitle
  
\setcounter{tocdepth}{2}

\setcounter{secnumdepth}{3}

\section{Introduction} \label{sec:intro}

\para{Formulations of general relativity and Higgs inflation.}

There are several formulations of general relativity, including the metric, the Palatini, the teleparallel and the symmetric teleparallel formulation, among others \cite{Einstein:1925, Einstein:1928a, Einstein:1928b, Einstein:1930, Krssak:2018ywd, Hehl:1976, Hehl:1978, Papapetrou:1978, Hehl:1981, Percacci:1991, Rovelli:1991, Nester:1998, Percacci:2009, Krasnov:2017, Gielen:2018, BeltranJimenez:2017, ferraris1982, BeltranJimenez:2019acz, Percacci:2020bzf}. They are based on different assumptions about spacetime degrees of freedom, in particular about the relation between the metric (or the tetrad) and the connection. When gravity is described by the Einstein--Hilbert action and matter does not couple directly to the connection, these formulations are equivalent. However, for more complicated gravitational actions \cite{Buchdahl:1960, Buchdahl:1970, ShahidSaless:1987, Flanagan:2003a, Flanagan:2003b, Sotiriou:2006, Sotiriou:2008, Olmo:2011, Borunda:2008, Querella:1999, Cotsakis:1997, Jarv:2018, Conroy:2017, Li:2007, Li:2008, Exirifard:2007, Enckell:2018b} or matter couplings \cite{Lindstrom:1976a, Lindstrom:1976b, Bergh:1981, Bauer:2008, Bauer:2010, Koivisto:2005, Rasanen:2017, Enckell:2018a, Markkanen:2017, Jarv:2017, Iosifidis:2018zwo, Rasanen:2018a, Rasanen:2018b, Aoki:2019rvi, Rubio:2019, Raatikainen:2019, Jinno:2019und, Shimada:2018, Iosifidis:2018jwu}, different formulations in general lead to different predictions.

Scalar fields couple directly to the connection via the Ricci scalar. Even if such a coupling is not included at tree level, it will be generated by quantum corrections \cite{Callan:1970}. Even if the coupling is put to zero on some scale, it runs and will thus be non-zero on other scales (although it can be negligible if the running is small). Thus, scalar fields break the equivalence between different formulations of general relativity. In particular, this is true for the Higgs field of the Standard Model of particle physics.

Inflation is the most successful scenario for the early universe, and it is typically driven by a scalar field \cite{Starobinsky:1980te, Kazanas:1980tx, Guth:1981, Sato:1981, Mukhanov:1981xt, Linde:1981mu, Albrecht:1982wi, Hawking:1981fz, Chibisov:1982nx, Hawking:1982cz, Guth:1982ec, Starobinsky:1982ee, Sasaki:1986hm, Mukhanov:1988jd}. The non-minimal coupling of the inflaton to gravity can leave an imprint on inflationary perturbations, so observations of the cosmic microwave background and large-scale structure may distinguish between different formulations of general relativity. If the Standard Model Higgs is the inflaton \cite{Bezrukov:2007} (for reviews, see \cite{Bezrukov:2013, Bezrukov:2015, Rubio:2018}; for an earlier similar model, see \cite{Futamase:1987, Salopek:1988}), it has (in the simplest cases) a large non-minimal coupling to the Ricci scalar, so different formulations can lead to large observational differences \cite{Bauer:2008, Rasanen:2017, Enckell:2018a, Markkanen:2017, Rasanen:2018a, Rasanen:2018b, Rubio:2019, Raatikainen:2019, Shaposhnikov:2020fdv, Tenkanen:2020dge}. Conclusions regarding perturbative unitarity, a key question on the particle physics side of Higgs inflation, can also be different \cite{Barbon:2009, Burgess:2009, Burgess:2010zq, Lerner:2009na, Lerner:2010mq, Hertzberg:2010, Bauer:2010, Bezrukov:2010, Bezrukov:2011a, Calmet:2013, Weenink:2010, Lerner:2011it, Prokopec:2012, Xianyu:2013, Prokopec:2014, Ren:2014, Escriva:2016cwl, Fumagalli:2017cdo, Gorbunov:2018llf, Ema:2019, McDonald:2020, Shaposhnikov:2020fdv, Enckell:2020lvn}. In the metric formulation, the scale of tree-level unitarity violation is naively below the inflationary scale. In the Palatini formulation, the scale of tree-level unitarity violation is shifted up, and it is possible that inflation may take place below this scale \cite{Bauer:2010, McDonald:2020, Shaposhnikov:2020fdv, Enckell:2020lvn}.

In addition to the non-minimal coupling of matter, a gravity sector more complicated than the Einstein--Hilbert action can lead to differences between formulations. One example is higher powers of the Ricci scalar, which have to be included because of loop corrections \cite{Barbon:2015, Salvio:2015kka, Salvio:2017oyf, Kaneda:2015jma, Calmet:2016fsr, Wang:2017fuy, Ema:2017rqn, Pi:2017gih, He:2018gyf, Gorbunov:2018llf, Ghilencea:2018rqg, Wang:2018kly, Gundhi:2018wyz, Karam:2018, Kubo:2018, Enckell:2018c, Ema:2019, Canko:2019mud, Enckell:2018a, Enckell:2018b, He:2020ivk, Bezrukov:2020txg}. Extended gravitational actions can also be motivated by top-down considerations involving more fundamental theories, such as loop quantum gravity (LQG).

\para{Loop quantum gravity.}

LQG is a candidate for a non-perturbative background-free theory of quantum gravity. Cosmology in LQG has often been studied in the loop quantum cosmology approach, which involves LQG-quantising\footnote{By LQG-quantising we mean that the background-free quantisation techniques used in the full LQG theory are mimicked as closely as possible. For example, the holonomy of the connection (rather than the connection) is quantised, the size of plaquettes is not shrunk to zero (as in the usual Wilson loop quantisation) because the minimal area eigenvalue is non-zero, the kinematical space is inequivalent to the one of Wheeler--de Witt quantisation but mimics that of full LQG, and so on.} a symmetry-reduced model (such as a Friedmann--Lema\^{\i}tre--Robertson--Walker spacetime) and studying the difference to the ordinary "von Neumann representation"-compatible quantisation \cite{Ashtekar:2011ni}. We instead consider the cosmological effects of new terms appearing in the LQG action to inflation at the classical level. (See also \cite{Bethke:2011ru, Bethke:2011kx} on inflationary gravitational waves in LQG.)

LQG comes in three main flavours: Hamiltonian form with the Ashtekar SL(2,C)-valued connection; Hamiltonian form with an SU(2)-valued connection; and covariant (or spin foam) form \cite{Rovelli:2004tv, Thiemann:2007zz, Gambini:2011zz, Rovelli:2014ssa}. In the first case the Hamiltonian is complex, so reality conditions have to be imposed to obtain real-valued geometry, in the second case the variables are real. In the Hamiltonian forms the action consists of the Einstein--Hilbert action (also called the Palatini action as the connection is an independent variable) plus the Holst action. The Holst term is the contraction of the Riemann tensor with the Levi--Civita tensor, multiplied by a constant whose inverse is called the Barbero--Immirzi parameter $\c$ \cite{Holst:1995pc}. As the Holst piece is of the same order in curvature as the Ricci scalar, it is not suppressed by an extra mass scale.

The choice $\c=\pm i$ gives the selfdual (or anti-selfdual) SL(2,C) action for LQG, for which all constraints are first class and can be solved \cite{Ashtekar:1986yd, Jacobson:1987qk}. However, as the action is complex, reality conditions have to be imposed, and it is not clear how to handle them when quantising. If $\c$ is real, we get the LQG action for the real-valued SU(2) connection, for which the Hamiltonian constraint is however complicated. In this case the spectrum of the area operator and the volume operator are discrete \cite{Bianchi:2010gc, Rovelli:1995ac}, unlike in the selfdual case when they are continuous \cite{Alexandrov:2001pa}.

The Holst term is central for black hole entropy. If the Barbero--Immirzi parameter is real, there are two possibilities depending on whether or not there is a chemical potential in the statistical treatment of the black hole entropy. (This depends on the quantisation of the dynamics, which is an open problem.) With no chemical potential, the entropy is inversely proportional to the Barbero--Immirzi parameter, and the semiclassical value of Bekenstein and Hawking \cite{Bekenstein:1972tm, Hawking:1974rv} is reproduced for $\c\approx0.274$ \cite{Ghosh:2004}. When a chemical potential is added, the correct semiclassical value can be obtained independent of the value of $\c$ \cite{Ghosh:2011fc, Bianchi:2012ui}. Black hole entropy is also independent of the Barbero--Immirzi parameter in the complex selfdual case \cite{Frodden:2012dq}.

In the case of pure gravity with the Einstein--Hilbert plus the Holst term, the Holst term does not contribute to the equations of motion at the classical level. The theory thus has a quantisation ambiguity as there is a one-parameter family of quantum theories corresponding to the classical theory. Another term sometimes considered in LQG is the topological Nieh--Yan invariant \cite{Nieh:1981, Calcagni:2009xz, Mercuri:2006um, Mercuri:2006wb, Mercuri:2009zt, Mercuri:2009zi, Date:2008rb}. Like the Holst term, it is dimension 2, and is hence not suppressed by a mass scale compared to the Ricci term. It is obvious that the Nieh--Yan term does not contribute at the classical level, as it is a total derivative. The case of the Holst term is more subtle. It vanishes when there is no torsion, and for minimally coupled matter, the equations of motion for the connection lead to the Levi--Civita connection, for which the torsion is zero\footnote{Assuming that non-metricity is zero, as usual in LQG.}.

When there is a source for torsion, the Holst term becomes dynamical. One case that has been studied in LQG is fermions whose kinetic term involves the spin connection \cite{Freidel:2005sn, Randono:2005, Perez:2005, Mercuri:2006um, Mercuri:2006wb, Bojowald:2007nu, Kazmierczak:2008}. Substituting the torsion generated by fermions back into the action leads to a four-fermion coupling that depends on the Barbero--Immirzi parameter, breaking the quantisation ambiguity.\footnote{Fermions can also be coupled to the Levi--Civita spin connection, so that they do not enter the connection equation of motion. Another possibility is to choose a modified kinetic term such that the dependence on the Barbero--Immirzi parameter disappears after solving the equations of motion \cite{Randono:2005, Freidel:2005sn, Mercuri:2006wb, Bojowald:2007nu}.}

Another possibility that has been considered is uplifting the Barbero--Immirzi parameter to a scalar field \cite{Castellani:1991et, Taveras:2008yf, Calcagni:2009xz, Cianfrani:2009sz, Mercuri:2009zi, Bombacigno:2016siz, TorresGomez:2008fj}, in which case a constant $\c$ is a low-energy approximation for when the field sits at the minimum. This Barbero--Immirzi field will source torsion. Substituting the torsion back into the action generates a free scalar field, and a potential would need to be added by hand for inflation. The coefficient of the Nieh--Yan term has likewise been promoted to a scalar field \cite{Mercuri:2009zi, Mercuri:2009zt, Calcagni:2009xz, Espiro:2014uda, Castillo-Felisola:2015ema, Cisterna:2018jsx}.

A third possibility that has been studied is that torsion is generated by the non-minimal coupling of a scalar field to the Ricci scalar \cite{Capovilla:1992tx, Montesinos:1999tq}. Such a coupling does not spoil the usual LQG quantisation procedure when $\c$ is real \cite{Montani:2009gn}. It has been considered both in loop quantum cosmology \cite{Bojowald:2006hd, Artymowski:2013qua, Artymowski:2012is} and from the perspective of black hole thermodynamics \cite{Ashtekar:2003jh, Ashtekar:2003zx}. Even if the Holst term is minimally coupled, the non-minimal coupling of the Ricci tensor will make it dynamical. If the non-minimally coupled field is the inflaton, the value of the Barbero--Immirzi parameter will be imprinted on the spectrum of perturbations produced during inflation. 

We consider non-minimal coupling of a scalar field to the Ricci scalar, Holst term and Nieh--Yan term during inflation, with particular attention to the Higgs case. Unlike for fermions, where the observational signature is negligible because the four-fermion interaction is suppressed by the Planck scale, we find that the scalar-generated torsion can have a significant effect, completely changing the inflationary predictions.

In \sec{sec:nm} we present the formalism, give the action where the Ricci scalar, Holst term and Nieh--Yan term are non-minimally coupled to a scalar field, solve the equation of motion of the connection and substitute back into the action. The physics of the non-minimal coupling is thus shifted to the kinetic term and potential of the scalar field. In \sec{sec:infl} we discuss inflationary behaviour in the case of Higgs inflation, including non-minimally coupled terms up to dimension 4, and in \sec{sec:conc} we summarise our results. We use the metric and the connection as the gravitational degrees of freedom, as these are more familiar to cosmologists. In appendix \ref{sec:tet} we present the calculation with tetrads, more familiar to people working on LQG.

\section{Non-minimal coupling to Ricci, Holst and Nieh--Yan terms} \label{sec:nm}

\subsection{Curvature, non-metricity and torsion}

We take the metric $g_{\a\b}$ and the connection $\Gamma^{\c}_{\a\b}$ to be independent degrees of freedom. The connection, defined with the covariant derivative as $\nabla_\b A^\a=\pat_\b A^\a + \Gamma^\a_{\b\c} A^\c$, can be decomposed as
\bea \label{Gamma}
  \Gamma^\c_{\a\b} &=& \mathring\Gamma^\c_{\a\b} + L^\c{}_{\a\b} = \mathring\Gamma^\c_{\a\b} + J^\c{}_{\a\b} + K^\c{}_{\a\b} \ ,
\eea
where $\mathring\Gamma_{\a\b}^\c$ is the Levi--Civita connection defined by the metric $g_{\a\b}$. As the difference of two connections, $L^\c{}_{\a\b}$ is a tensor, known as the distortion. In the second equality we have decomposed it into the disformation $J_{\a\b\c}$ and the contortion $K_{\a\b\c}$, defined as
\bea \label{L}
  J_{\a\b\c} &\equiv& \frac{1}{2} \left( Q_{\a\b\c}  - Q_{\c\a\b} - Q_{\b\a\c} \right) \ , \qquad K_{\a\b\c} \equiv \ha ( T_{\a\b\c} + T_{\c\a\b} + T_{\b\a\c} ) \ ,
\eea
where $Q_{\a\b\c}$ and $T_{\a\b\c}$ are the non-metricity and the torsion, respectively, defined as
 \bea \label{TQ}
  \qquad Q_{\c\a\b} \equiv \nabla_\c g_{\a\b}  \ , \qquad T^\c{}_{\a\b} &\equiv& 2 \Gamma^{\c}_{[\a\b]} \ .
\eea
Note that $Q_{\c\a\b}=Q_{\c(\a\b)}$, $\nabla_\c g^{\a\b}=-Q_{\c}^{\ \a\b}$, $J_{\a\b\c}=J_{\a(\b\c)}$ and $K^\c{}_\a{}^\b=K^{[\c}{}_\a{}^{\b]}$.

The two non-metricity vectors are defined as
\bea \label{Qvec}
  Q^\c \equiv g_{\a\b} Q^{\c\a\b} \ , \qquad \hat Q^\b \equiv g_{\a\c} Q^{\a\b\c} \ ,
\eea
and the torsion vector and torsion axial vector\footnote{Despite the name, the parity transformation properties of the torsion vector and axial vector are not necessarily those of a vector and pseudovector. How they transform depends on the solution for the torsion tensor.} are defined as, respectively,
\bea \label{Tvec}
  T^\b \equiv g_{\a\c} T^{\a\b\c} \ , \qquad \hat T^\a \equiv \frac{1}{6} \epsilon^{\a\b\c\d} T_{\b\c\d} \ ,
\eea
where $\epsilon_{\a\b\c\d}$ is the Levi--Civita tensor. Note that $\nabla_\a\sqrt{-g}=\ha \sqrt{-g} Q_\a$.

The Riemann tensor can be decomposed into the Levi--Civita and the distortion contribution as
\bea \label{Riemann}
  R^{\a}{}_{\b\c\d} = \mathring R^{\a}{}_{\b\c\d} + 2 \mathring \nabla_{[\c} L^\a{}_{\d]\b} + 2 L^\a{}_{[\c|\mu|} L^\mu{}_{\d]\b} \ ,
\eea
where $\mathring{}$ denotes a quantity defined with the Levi--Civita connection. The curvature $R^{\a}{}_{\b\c\d}$, non-metricity $Q_{\a\b\c}$ and torsion $T_{\a\b\c}$ are the complete set of tensors that characterise the geometry of a manifold.

There are exactly two geometrical scalars that are linear in the Riemann tensor \re{Riemann} and quadratic in the connection: the Ricci scalar and the Holst term. They are defined as, respectively,
\bea \label{RhatR}
  R &\equiv& \delta_\a{}^\c g^{\b\d} R^{\a}{}_{\b\c\d} = \mathring R + Q + T + \mathring\nabla_\a ( Q^\a - \hat Q^\a + 2 T^\a ) - T_\a ( Q^\a - \hat Q^\a ) + Q_{\a\b\c} T^{\c\a\b} \el
  \hat R &\equiv& \ha g_{\a\mu} \epsilon^{\mu\b\c\d} R^\a{}_{\b\c\d} = - 3 \mathring\nabla_\a \hat T^\a + \frac{1}{4} \epsilon^{\a\b\c\d} T_{\mu\a\b} T^\mu{}_{\c\d} + \ha \epsilon^{\a\b\c\d} Q_{\a\b\mu} T^\mu{}_{\c\d} \ ,
\eea
where we have used \re{L}--\re{Riemann} to separate the contributions of curvature, non-metricity and torsion. The non-metricity scalar and the torsion scalar are defined as $Q \equiv \frac{1}{4} Q_{\a\b\c} Q^{\a\b\c} - \frac{1}{2} Q_{\a\b\c} Q^{\c\a\b} - \frac{1}{4} Q_\a Q^\a + \frac{1}{2} Q_\a \hat Q^\a$ and $T \equiv \frac{1}{4} T_{\a\b\c} T^{\a\b\c} - \frac{1}{2} T_{\a\b\c} T^{\c\a\b} - T_\a T^\a$, respectively. We also consider the Nieh--Yan term $\mathring\nabla_\a \hat T^\a$, which is equivalent to the Holst term plus a term quadratic in the torsion and a term involving non-metricity. In LQG non-metricity is usually taken to be zero a priori, so this term is absent. Although we are motivated by LQG, our approach is bottom-up, so we keep the non-metricity (although it will turn out it can be set to zero without loss of generality).

\subsection{The action and the connection}

We consider an action with the Ricci scalar, the Holst term and the Nieh--Yan term coupled to a scalar field $h$, which we will later identify with the Standard Model Higgs,
\bea
  \label{actionJ} \!\!\!\!\!\!\!\!\!\!\!\!\! S &=& \int\rmd^4 x \sqrt{-g} \left[ \ha F(h) R + \ha H(h) \hat R + \frac{3}{2} Y(h) \mathring\nabla_\a \hat T^\a - \ha K(h) g^{\a\b} \pat_\a h \pat_\b h - V(h) \right] \el
  &=& \int\rmd^4 x \sqrt{-g} \left[ \ha F(h) R + \ha H(h) \hat R - \frac{3}{2} \hat T^\a \partial_\a Y(h) - \ha K(h) g^{\a\b} \pat_\a h \pat_\b h - V(h) \right] \ ,
\eea
where on the second line we have discarded a boundary term. We neglect fermions. The coefficients have been chosen such that for $Y=H$, the $\mathring\nabla_\a \hat T^\a$ parts in the Holst term and the Nieh--Yan term cancel each other. Note the similarity of the Nieh--Yan term to the torsion vector coupling that appears in the teleparallel formulation \cite{Raatikainen:2019}. The usual LQG case corresponds to $F=1$, $H=1/\c$, $Y=0$, or alternatively $F=1$, $H=0$, $Y=1/\c$, where $\c$ is the Barbero--Immirzi parameter. (We choose units such that the Planck scale is unity.)

Varying \re{actionJ} with respect to the connection $\Gamma^\c_{\a\b}$, we get the equation of motion
\bea \label{Gammaeq}
  && - F Q_{\c\a\b} + F \hat Q_\b g_{\a\c} - H \epsilon_{\a\b}{}^{\mu\nu} Q_{\mu\nu\c} + F g_{\a[\b} ( Q_{\c]} + 2 T_{\c]} ) \el
  && + F T_{\a\b\c} + H \epsilon_{\a\b\c}{}^{\mu} T_\mu + \ha H \epsilon_{\a\b}{}^{\mu\nu} T_{\c\mu\nu} = - 2 g_{\a [\b} \pat_{\c]} F - \epsilon_{\a\b\c}{}^{\mu} \pat_\mu ( H - Y ) \ .
\eea
The general solution of \re{Gammaeq} has the form
\bea \label{Gammasol}
  Q_{\c\a\b} &=& q_1(h) g_{\a\b} \pat_\c h + 2 q_2(h) g_{\c(\a} \pat_{\b)} h \el
  T_{\a\b\c} &=& 2 t_1(h) g_{\a[\b} \pat_{\c]} h + t_2(h) \epsilon_{\a\b\c}{}^{\mu} \pat_\mu h \ . 
\eea
The definitions \re{Qvec} and \re{Tvec} give
\bea \label{QTsol}
  Q_\a &=&  ( 4 q_1 + 2 q_2 ) \pat_\a h \ , \quad \hat Q_\a = ( q_1 + 5 q_2 ) \pat_\a h \el
  T_\a &=& -3 t_1 \pat_\a h \ , \quad\quad\quad\quad \hat T_\a = t_2 \pat_\a h \ .
\eea
As non-metricity and torsion are only sourced by the scalar field, they can be written in terms of gradients of the scalar field, and reduce to the four vectors \re{QTsol}.

Inserting \re{Gammasol} into \re{Gammaeq}, we get
\bea \label{qt}
  q_2 &=& 0 \el
  2 t_1- q_1 &=& \frac{F F' + H ( H ' - Y' )}{F^2 + H^2} \el
  t_2 &=& \frac{H F' - F ( H' - Y' )}{F^2 + H^2} \ ,
\eea
where prime denotes derivative with respect to $h$. If $F=H$, the action has the extra symmetry of invariance under the duality transformation $R_{\a\b\c\d}\to\ha\epsilon_{\c\d}{}^{\a\b}R_{\a\b\a\b}$, which maps $R\leftrightarrow\hat R$. Then the Holst term does not contribute to the equations of motion if $Y'=0$, as is easily seen by a conformal transformation to the Einstein frame. If $Y'\neq0$, the Holst term with $H=F$ simply effectively shifts $Y\to Y/2$.

The equations of motion do not fix $q_1$ and $t_1$ separately, only the combination $2 t_1-q_1$. This well-known feature is due to invariance of the action \re{actionJ} under the projective transformation $\Gamma^\c_{\a\b}\to\Gamma^\c_{\a\b}+\delta^\c{}_\b A_\a$, where $A_\a$ is an arbitrary vector \cite{Hehl:1978}. The Riemann tensor transforms as $R_{\a\b\c\d}\to R_{\a\b\c\d}+ g_{\a\b} ( 2 \nabla_{[\c} A_{\d]} + T^\mu{}_{\c\d} A_\mu )$, so the Ricci scalar is invariant due to the symmetry of its contraction, and the Holst term is invariant due to the antisymmetry of its contraction. The Nieh--Yan term is invariant because $\hat T^\a$ is invariant. If $A_\a=\partial_\a A$ for some scalar $A$, the non-metricity and the torsion transform as $q_1\to q_1-2 A$, $t_1\to t_1-A$, $q_2\to q_2$, $t_2\to t_2$. 

The projective symmetry can be explicitly broken in the action \cite{Rasanen:2018b, Shimada:2018, Aoki:2019rvi, BeltranJimenez:2019acz, Iosifidis:2018jwu}. Short of that, the projective invariance is often fixed in the Palatini formulation by assuming a priori that the connection is symmetric, $T_{\a\b\c}=0$. When the Holst term is included, \re{qt} shows that this is not possible unless $F'/F=(H'-Y')/H$. (Note the similarity of this condition to the condition for the torsion vector coupling to vanish in the teleparallel formulation \cite{Raatikainen:2019}.) In the tetrad formulation used in LQG, it is instead commonly assumed that the covariant derivative of the tetrad is zero, which goes under the name tetrad postulate, meaning $Q_{\a\b\c}=0$. The scalar field kinetic term and potential are trivially invariant under the projective transformation as they do not depend on the connection. When deriving the equation of motion for the scalar field, the requirement that a total covariant derivative of a scalar field term reduces to a boundary term picks out the Levi--Civita connection, so the full connection does not appear in the scalar field equation of motion.

Following the LQG convention, we fix the projective symmetry by setting $q_1=0$, so non-metricity is zero. In fact, we could have put $Q_{\a\b\c}=0$ from the beginning, as the following reasoning shows. We can get rid of the non-minimal coupling $F$ to $R$ by a conformal transformation. As a conformal transformation (see \re{confQ} below) can only change $q_2$, not $q_1$, $F$ cannot generate a $q_2$ term. And as we can perform a conformal transformation to cancel the source term involving the Holst or the Nieh--Yan term, they also cannot generate $q_2$. And $q_1$ can always be transformed into $t_1$ by the projective transformation. For a different action, setting $Q_{\a\b\c}=0$ may involve loss of generality \cite{Rasanen:2018b}.

\subsection{Einstein frame action}

The coupled equations of motion for the scalar field, metric and connection can be simplified by choosing suitable coordinates in field space. If we make the conformal transformation
\bea \label{conf}
  g_{\a\b} &\to& \Omega(h)^{-1} g_{\a\b}
\eea
and absorb the changes in the functions of $h$ in the action, they transform as
\bea \label{fung}
  F &\to& \Omega^{-1} F \el
  H &\to& \Omega^{-1} H \el
  \partial_\a Y &\to& \Omega^{-1}  \partial_\a Y \el
  K &\to& \Omega^{-1} K \el
  V &\to& \Omega^{-2} V \ ,
\eea
and the non-metricity transforms as
\bea \label{confQ}
  Q_{\c\a\b} = \nabla_\c g_{\a\b} &\to& \Omega^{-1} ( \nabla_\c g_{\a\b} - g_{\a\b} \pat_\c \ln \Omega ) \ .
\eea

We choose field coordinates where the Ricci scalar is minimally coupled to the scalar field (\ie the Einstein frame), which corresponds to $\Omega=F$. The action then reads
\bea
  \label{actionE} S &=& \int\rmd^4 x \sqrt{-g} \left[ \ha R + \ha \frac{H(h)}{F(h)} \hat R - \frac{3}{2} \hat T^\a \frac{\partial_\a Y(h)}{F(h)} - \ha \frac{K(h)}{F(h)} g^{\a\b} \pat_\a h \pat_\b h - U(h) \right] \ ,
\eea
where we have denoted $U\equiv V/F^2$.

Inserting the connection \re{Gammasol} (with $F\to1$, $H\to H/F$ and $Y'\to Y'/F$) back into the action, decomposing $R$ and $\hat R$ into their Levi--Civita, non-metricity and torsion parts with \re{RhatR}, setting the non-metricity to zero and inserting the torsion \re{qt}, we get (dropping a boundary term)
\bea
  \label{actionG} S &=& \int\rmd^4 x \sqrt{-g} \bigg[ \ha \mathring R - \ha \left\{ \frac{K}{F} + 6 t_1^2 - \frac{3}{2} t_2^2 - 3 t_2 [ (H/F)' - Y'/F- 2 t_1 H/F ] \right\} g^{\a\b} \pat_\a h \pat_\b h \el
  && - U \bigg] \ .
\eea
Inserting $t_1$ and $t_2$ from \re{qt}, we arrive at the simple expression
\bea
  \label{actionEH} S &=& \int\rmd^4 x \sqrt{-g} \bigg[ \ha \mathring R - \ha \bigg\{ \frac{K}{F} + \frac{3}{2} \frac{ F^2 + H^2 }{ F^2 } t_2^2 \bigg\} g^{\a\b} \pat_\a h \pat_\b h - U \bigg] \el
  &=& \int\rmd^4 x \sqrt{-g} \bigg[ \ha \mathring R - \ha \bigg\{ \frac{K}{F} + \frac{3}{2} \frac{ [ (H/F)'-Y'/F ]^2 }{(H/F)^2+1} \bigg\} g^{\a\b} \pat_\a h \pat_\b h - U \bigg] \el
  \label{actionh1} &=& \int\rmd^4 x \sqrt{-g} \bigg[ \ha \mathring R - \ha \bigg\{ \frac{K}{F} + \frac{3}{2} \frac{[ H F' - F ( H' - Y' ) ]^2}{ F^2 ( F^2+H^2 ) } \bigg\} g^{\a\b} \pat_\a h \pat_\b h - U \bigg] \el
  \label{actionh} &\equiv& \int\rmd^4 x \sqrt{-g} \left[ \ha \mathring R - \ha \tilde K(h) g^{\a\b} \pat_\a h \pat_\b h - U(h) \right] \ .
\eea
The geometrical contribution of the torsion has been shifted to the scalar kinetic term and the $1/F^2$ modification of the potential, and only the Levi--Civita connection appears. When we vary this action with respect to $g_{\a\b}$ and $h$, we get equations of motion that are equivalent to those of the original action \re{actionJ}, which has a non-trivial gravity part (and hence connection). There is one subtle difference: varying the Einstein frame Levi--Civita action \re{actionEH} leads to boundary terms that depend on the derivative of the variation of the metric. In order to derive the equations of motion, we need to include the York--Gibbons--Hawking boundary term \cite{York:1972, Gibbons:1977} to cancel this contribution. In the original action \re{actionJ}, there is no such problem, as the variation of the connection can be taken to vanish on the boundary independently of the metric. From the Palatini perspective, having to add a boundary term to the Einstein--Hilbert action is an artifact of solving part of the equations of motion and inserting the result back into the action. (We discarded boundary terms in the derivation.)

\subsection{Recovering the metric and Palatini cases} \label{sec:rec}

The action \re{actionh} reduces to the well-known Palatini case with a non-minimal coupling only to the Ricci scalar \cite{Bauer:2008} when $Y'=F (H/F)'$. Apart from the trivial case $H=Y=0$, this also happens when $Y=0$, $H=\a F$, where $\a$ is an arbitrary constant. If both the Holst and the Nieh--Yan term are non-zero, the condition means that their derivative parts cancel in the action, leaving only a quadratic torsion term.

The results of the metric formulation with a non-minimal coupling to the Ricci scalar are recovered when
\bea \label{limit}
  F' = ( Y' - H' )( H/F \pm \sqrt{ (H/F)^2 + 1 } ) \ .
\eea
A particularly simple case is $H=0$, $Y=\pm F$, when there is no Holst term and the coupling functions of the Ricci term and the Nieh--Yan term are identical (possibly up to a sign). Another possibility is $Y=0$, $F=\pm\a^{-1}\sqrt{1+2\a H}$, where $\a$ is an arbitrary non-zero constant.

\section{Inflation} \label{sec:infl}

\subsection{The coupling functions and the potential} \label{sec:spec}

Let us now discuss inflation with the Standard Model Higgs. A field-dependent kinetic term corresponds to a monotonic field-dependent remapping of the potential. Including only terms of up to dimension 4 in the action \re{actionJ} and taking into account that only even powers of the field appear, we have (note that $Y$ is defined only up to an additive constant)
\bea \label{Higgs}
  K = K_0 \ , \quad F = F_0  ( 1 + \xi h^2 ) \ , \quad H = F_0 ( H_0 + H_1 h^2 ) \ , \quad Y = F_0 Y_1 h^2 \ ,
\eea
where $K_0, F_0, \xi, H_0, H_1$ and $Y_1$ are constants. In LQG with the Holst term, $H_0=1/\c$. Observational limits on these couplings are very weak, as they effectively only modify the Higgs potential for large field values. In the metric formulation, collider measurements give $|F_0\xi|<2.6\times10^{15}$ \cite{Atkins:2012}. In the Palatini case, non-inflationary constraints on the non-minimal couplings are likewise expected to be so high as not to affect our analysis.

The kinetic function defined in \re{actionh} is
\bea \label{K}
  \tilde K &=& \frac{K}{F} + \frac{3}{2} \frac{[ H F' - F ( H' - Y' ) ]^2}{ F^2 ( F^2+H^2 ) } \el
  &=& \frac{K_0}{F_0 ( 1 + \xi h^2 ) } + 6 h^2 \frac{ ( Y_1 - H_1 + H_0 \xi + Y_1 \xi h^2)^2 }{ ( 1 + \xi h^2 )^2 [ 1 + H_0^2 + 2 ( H_0 H_1 + \xi ) h^2 + ( H_1^2 + \xi^2 ) h^4 ] } \ .
\eea
The kinetic function and thus the physics is invariant under the simultaneous sign change of $H_0$, $H_1$ and $Y_1$. In the small field limit $h\ll1$ the second term falls off like $h^2$, and the canonically normalised field is $\chi=\sqrt{K_0/F_0} h$. In general, the transformation between $h$ and the canonical field $\chi$ is
\bea
  \frac{\rmd\chi}{\rmd h} = \pm \sqrt{ \tilde K } \ .
\eea

We consider the Higgs tree-level potential, so
\bea \label{U}
  U(\chi) = \frac{\lambda}{4 F_0^2} \frac{ [ h(\chi)^2-v^2 ]^2}{ [ 1 + \xi h(\chi)^2 ]^2 } \ ,
\eea
where $\lambda$ and $v$ are constants. The constants $K_0$ and $F_0$ effectively rescale the values of $\lambda$ and $v$ when we consider the potential in terms of the canonically normalised field \cite{Rasanen:2018b}, and we henceforth take $K_0=F_0=1$. The quartic coupling $\l$ has the value $0.13$ at the electroweak scale, and without the non-minimal gravitational couplings it runs down with increasing field value. The running depends on the electroweak scale values of the Higgs mass, top quark mass, and QCD coupling constant. For the measured mean values, $\l$ crosses zero around $10^{11}$ GeV $\sim10^{-7}$, in the case all when non-minimal couplings are zero \cite{Espinosa:2015a, Espinosa:2015b, Iacobellis:2016, Espinosa:2016nld, Hoang:2020iah}. The running is highly sensitive to the input electroweak scale values, and positivity of $\l$ up to the Planck scale is within the 2$\sigma$ limits \cite{Espinosa:2015a, Espinosa:2015b, Iacobellis:2016, Espinosa:2016nld, Hoang:2020iah}. The non-minimal couplings we consider can also change the renormalisation group running. We do not consider running, and take $\l$ at the inflationary scale to be a free positive parameter, limited by $\l<0.1$ to avoid strong coupling. If we used $\l<0.01$ instead, the upper limit for $\xi$ we find would decrease by one order of magnitude, which in the case $\xi>0$ correspondingly brings the lower limit for $r$ up by one order of magnitude.

The first slow-roll parameters are
\bea \label{SR}
  \epsilon &=& \ha \left( \frac{U'}{U} \right)^2 \ , \qquad \eta = \frac{U''}{U} \ , \qquad \sigma_2 = \frac{U'}{U} \frac{U'''}{U} \ , \qquad \sigma_3 = \left( \frac{U'}{U} \right)^2 \frac{U''''}{U} \ ,
\eea
where prime denotes derivative with respect to $\chi$. 

The amplitude, spectral index, running, running of the running of the scalar perturbations, and the tensor-to-scalar ratio are, respectively,
\bea
  \label{A} A_s &=& \frac{1}{24\pi^2} \frac{U}{\epsilon} = 2.099 \, e^{\pm 0.014} \, 10^{-9} \\
  \label{n} n_s &=& 1 - 6 \epsilon + 2 \eta = 0.9625 \pm 0.0048 \\
  \label{a} \a_s &=& - 24 \epsilon^2 + 16 \epsilon \eta - 2 \sigma_2 = 0.002 \pm 0.010 \\
  \label{b} \b_s &=& - 192 \epsilon^3 + 192 \epsilon^2 \eta - 32 \epsilon \eta^2 - 24 \epsilon \sigma_2 + 2 \eta \sigma_2 + 2 \sigma_3 = 0.010 \pm 0.013 \\
  \label{r} r &=& 16 \epsilon < 0.067 \ ,
\eea
where the observational values with 68\% C.L. limits are from Planck and BICEP2/Keck cosmic microwave background (CMB) data at the pivot scale 0.05 Mpc$^{-1}$ \cite{Planck2018}. The value for $r$ assumes zero running of the running. The number of e-folds until the end of inflation is
\bea \label{N}
  N=\int_{\chi_\textrm{end}}^\chi \frac{\rmd\chi}{\sqrt{2\epsilon}} \ ,
\eea
where $\chi_\textrm{end}$ is the field value at the end of inflation (approximating that the field is in slow-roll until the end of inflation). The number of e-folds at the pivot scale is
\begin{equation} \label{Nval}
  N = 56 - \Delta N - \frac{1}{4} \ln \frac{0.067}{r} \ ,
\end{equation}
where $\Delta N$ accounts for the effect of reheating. Reheating is sensitive to the shape of the potential. With a non-minimal coupling only to the Ricci scalar, in the Palatini formulation with a tree-level potential the reheating is almost instant, $\Delta N\ll1$ \cite{Rubio:2019}. In the metric case it is not clear whether $\Delta N=4$ or $\Delta N\ll1$ \cite{Bezrukov:2008ut, GarciaBellido:2008ab, Figueroa:2009, Figueroa:2015, Repond:2016, Ema:2016, DeCross:2016, Sfakianakis:2018lzf, Hamada:2020kuy}. We assume instant reheating.

\subsection{Plateau inflation}

Let us first consider inflation on the asymptotically flat plateau, which the potential has when $\xi>0$. When the Holst and the Nieh--Yan term are zero, this is the only inflationary regime. With either or both non-zero, plateau inflation remains qualitatively the same, and the first slow-roll observables in terms of the number of e-folds are (see \eg \cite{Rasanen:2017} for details)
\bea \label{plateau}
  A_s &=& \frac{N^2}{12 \pi^2} \frac{\lambda}{ \xi + \frac{6 \xi^2 Y_1^2}{H_1^2+\xi^2} } \el
  n_s &=& 1 - \frac{2}{N} - \frac{3 r}{16} \el
  r &=& \frac{2}{N^2} \left( \frac{1}{\xi} + \frac{ 6 Y_1^2 }{ H_1^2 + \xi^2 } \right) = \frac{\lambda}{6\pi^2 A_s \xi^2} \ .
\eea
The term $\frac{3 r}{16}$ in the expression for $n_s$ has sometimes been dropped. While it is negligible for small $r$, for the maximum value $r=0.067$ it gives a correction of $-0.012$. For $N=56$ \cite{Rubio:2019}, we get $n_s=0.96-\frac{3 r}{16}$, in agreement with observations. In contrast to the cases $H=Y=0$, the amplitude $A_s$ can be small without a large $\xi$, if the Nieh--Yan term coupling $Y_1$ is large instead. However, the observational upper limit \re{r} on $r$ combined with the value \re{A} of $A_s$ anyway requires $\xi>10^4\sqrt{\lambda}$, so unless $\lambda\ll1$, we have $\xi\gg1$. The tensor-to-scalar ratio $r$ can be adjusted up or down from the metric case result $12/N^2$ by shifting the parameters. The minimum value is $r=5\times10^{-13}$ (assuming $\lambda<0.1$), corresponding to the tree-level Palatini case with a non-minimal coupling only to the Ricci scalar.

If the Holst term is zero, only plateau inflation is possible. In this case the behaviour is identical to the teleparallel case studied in \cite{Raatikainen:2019}. However, if $H_1\neq0$, we can get qualitatively different inflationary behaviour. Let us first look at some interesting subcases. We have verified all results by numerically scanning the parameter space.

\subsection{$Y=0$} \label{sec:Y0}

Let us consider the case when the Nieh--Yan term is zero, but not the Holst term. We see from \re{plateau} that the Holst term plays no role in plateau inflation, unless its coupling is large. This is because the Holst contribution to the kinetic function \re{K} decreases like $1/h^6$ for large $h$, in contrast to the $1/h^2$ suppression of the $\xi$ term. So even though the Holst term is non-zero because $F$ generates torsion, its numerical contribution is negligible. In particular, this is the case if we take $H_0=1/\c\approx3.6$, where $\c=0.274$ is the value determined from black hole entropy in LQG without chemical potential \cite{Ghosh:2004}. If the Holst term coupling is large, $n_s$ can be shifted down on the plateau.

However, if $H_0$ is much larger than $\xi$ and $H_1$, there is another inflationary regime in addition to plateau inflation. The contribution of the Holst term can dominate the kinetic function \re{K} in an intermediate regime even though it is subleading in the limit $h\to\infty$. When $H_0$ dominates over all other terms and $|\xi| h^2\gg1$, the kinetic term is $\tilde K\simeq6/h^2$. This agrees with the metric formulation plateau case \cite{Bezrukov:2007}, giving $n_s=1-2/N=0.96$ and $r=12/N^2=4\times10^{-3}$ for $N=56$. However, now this solution also exists if $\xi<0$. If the other terms also contribute, the results for $r$ remain the same, but $n_s$ can be adjusted downwards. The running parameters $\a$ and $\b$ can also take a range of values outside those of plateau inflation driven by a non-minimal coupling to the Ricci scalar. In this inflationary regime, $h$ at the pivot scale can be as small as $2\times10^{-3}$, in contrast to usual plateau inflation, where $h\approx0.08$ in the metric formulation and $h\approx20$ in the Palatini formulation. Interestingly, this case is possible even if $H_1=0$, \ie if the Holst coupling is constant. The non-minimal coupling $F$ generates torsion, making the Holst term dynamical, its effect enhanced by the large value of $H_0$.

If $\xi=0$, there is also a third inflationary regime, which gives predictions close to the metric case, as we discuss in the next section.

\subsection{$\xi=0$} \label{sec:xi0}

If the non-minimal coupling to the Ricci scalar is zero, the potential is not asymptotically flat. Nevertheless, we can have an intermediate flat regime where inflation can be successful (meaning the predictions agree with observations). The kinetic function \re{K} simplifies to
\bea \label{Kxi0}
  \tilde K &=& 1 + \frac{3}{2} \frac{ ( H' - Y' )^2}{ 1 + H^2 } \el
  &=& 1 + 6 h^2 \frac{ (Y_1 - H_1)^2 }{ 1 + ( H_0 + H_1 h^2 )^2 } \ .
\eea
We take $H_0>0$. (The case $H_0=0$ does not lead to successful inflation, and negative values of $H_0$ are related by symmetry to positive values.) For successful inflation, the second term has to dominate, in which case the canonical field is
\bea
  \chi &=& \int \rmd h \sqrt{\tilde K} \simeq \sqrt{\frac{3}{2}} \left|\frac{H_1-Y_1}{H_1} \right| \textrm{arsinh} ( H_0 + H_1 h^2 ) + \chi_0 \ ,
\eea
which gives
\bea
  h^2 &=& H_1^{-1} \sinh[ \sqrt{\frac{2}{3}} \Delta ( \chi + \chi_0 ) ] - H_1^{-1} H_0 \ ,
\eea
where $\Delta\equiv\left|\frac{H_1}{H_1-Y_1}\right|$ and $\chi_0\equiv\sqrt{\frac{3}{2}} \Delta^{-1} \textrm{arsinh} H_0$, so that $\chi=0$ corresponds to $h=0$. The potential \re{U} reads
\bea
  U &=& \frac{\lambda}{4 H_1^2} \left\{ \sinh\left[ \sqrt{\frac{2}{3}} \Delta ( \chi + \chi_0 ) \right] - H_0 \right\}^2 \ .
\eea
In the limit $\sinh[\sqrt{\frac{2}{3}}\Delta(\chi+\chi_0)]\gtrsim1$ (which is required for inflation satisfying the observational constraints \re{A}--\re{r} and the constraint on the number of e-folds), the amplitude, spectral index, tensor-to-scalar ratio and the number of e-folds from \re{A}--\re{N} are ($\simeq$ indicates dropping corrections of order $1/\sinh[\sqrt{\frac{2}{3}}\Delta(\chi+\chi_0)]^2$)
\bea \label{xi0results}
  A_s &\simeq& \frac{\lambda H_0^2}{128 \pi^2 \Delta^2 H_1^2} \frac{x^4}{(1+x)^2} \el
  n_s &=& 1 - \frac{8 \Delta^2}{3 x} - \frac{r}{4} \el
  r &\simeq& \frac{64 \Delta^2}{3 x^2} \el
  N &\simeq& \frac{3}{4\Delta^2} [ x - \ln(1+x) ] \ ,
\eea
where we have denoted $1+x\equiv H_0/\sinh[\sqrt{\frac{2}{3}}\Delta(\chi+\chi_0)]$.

The expressions for the running and the running of the running are also straightforward to write down; they are within the observational ranges \re{a}-\re{b}. In the limit $x\gg1$ we  can drop the logarithmic corrections to get
\bea
  A_s &\simeq& \frac{\lambda \Delta^2 H_0^2}{72 \pi^2 H_1^2} N^2 \el
  n_s &\simeq& 1 - \frac{2}{N} - \frac{r}{4} \el
  r &\simeq& \frac{12}{\Delta^2 N^2} \simeq \frac{\lambda H_0^2}{6\pi^2 A_s H_1^2} \ .
\eea
These equations are almost identical to those in the plateau inflation case with the replacement $\Delta^{-2}\to \frac{1}{6 \xi} + \frac{Y_1^2}{H_1^2+\xi^2}$, $\frac{H_1^2}{H_0^2}\to\xi^2$. The tensor-to-scalar ratio can be as large as the observational upper limit and as small as desired. The only difference is the last term for $n_s$ is $-\frac{r}{4}$ instead of $-\frac{3 r}{16}$, but the difference is $4\times10^{-3}$ even for the maximum observationally allowed value of $r$.

In the pure Holst case, $Y_1=0$, we have $\Delta=1$, and the predictions are identical to the metric plateau case, as mentioned above. This can be seen from \re{limit}: if $Y'=F'=0$ and $H\gg F$, the action is the same as in the metric case.

In the pure Nieh--Yan case, $\xi=H_0=H_1=0$, there are no inflationary solutions that agree with observations. (The case with $\xi=H_1=0$ but $H_0\neq0$ is equivalent to this case with the change $Y_1^2/(1+H_0^2)\to Y_1^2$.) In this case the kinetic function \re{Kxi0} grows like $h^2$ for large $h$, mapping the potential $\frac{1}{4} \lambda h^4$ to the potential $\frac{\lambda}{6 Y_1} \chi^2$ at large field values. Adjusting $Y_1$ interpolates between the quartic and the quadratic potential. While the spectral index of the quadratic potential (unlike the quartic potential) agrees with the data, the tensor-to-scalar ratio $r$ is too large in both cases \cite{Planck2018}. (A similar situation arises in the teleparallel formulation \cite{Raatikainen:2019}.) The value of $r$ can be decreased by including a $R^2$ term in the action \cite{Enckell:2018b}.

\subsection{$\xi<0$} \label{sec:xineg}

Including all three coupling terms (to the Ricci scalar, the Holst term and the Nieh--Yan term) makes it possible to have inflationary behaviour beyond the plateau and the above subcases. Let us first discuss the case $\xi<0$.

Given that we can have successful inflation when $\xi=0$, by continuity we expect this to be possible also for small negative values of $\xi$. However, there are also successful inflationary models for large negative values of $\xi$. If $\xi<0$, the non-minimal coupling $F$ goes to zero at $h=1/\sqrt{|\xi|}$. The kinetic function \re{K} correspondingly diverges, so we have an $\a$-attractor \cite{Ferrara:2013}, found for Higgs inflation for another action in \cite{Rasanen:2018b}. (Plateau Higgs inflation can also be viewed in terms of an $\a$-attractor \cite{Galante:2014, Rubio:2018, Karananas:2020qkp}.) However, the $\a$-attractor behaviour in the limit $F\to0$ does not give successful inflation, as $n_s$ and/or $r$ are wrong. Nevertheless, there are other kinds of successful inflationary models with $\xi<0$.

In the case $H_0=0$ there are no viable inflationary models. In the case $Y=0$ there are no viable models if also $H_1=0$. If $Y=0$ and $H_1\neq0$, the only viable case is the one discussed in \sec{sec:Y0}. If we allow both $H_0\neq0$ and $Y\neq0$, the range of predictions widens.

One particular new case is inflection point inflation. At an inflection point $\eta=0$, so the spectral index there is $n_s=1-3r/8$. The observational limit $r<0.067$ in \re{r} then gives $n_s>0.97$, which is at the upper end of the observational range \re{n}. So if there is an inflection point close to the pivot scale, the amplitude of inflationary gravitational waves is close to the observational upper limit. This is the reason an inflection point due to quantum corrections \cite{Allison:2013uaa, Bezrukov:2014bra, Hamada:2014iga, Bezrukov:2014ipa, Rubio:2015zia, Fumagalli:2016lls, Enckell:2016xse, Bezrukov:2017dyv, Rasanen:2017, Masina:2018ejw, Salvio:2017oyf, Ezquiaga:2017fvi, Rasanen:2018a} was highlighted after the claimed detection of gravitational waves by the BICEP2 instrument (which turned out to be incorrect). We find models with an inflection point exactly at the pivot scale that agree with observations, apart from this tension. Inflection point in Higgs inflation from classical contributions to the action that can generate torsion has been earlier discussed in \cite{Rasanen:2018b}.

We scanned numerically over the five-dimensional parameter space ($h$, $\xi$, $H_0$, $H_1$, $Y_1$) with an adaptive Monte Carlo method. We take the range $[-10^{10}, 0]$ for $\xi$, $[-10^{10}, 10^{10}]$ for $H_0$ and $H_1$, and $[0, 10^{10}]$ for $h$ and $Y_1$. (We can fix one sign among $H_0$, $H_1$ and $Y_1$ without affecting the physics.) We check that observables at the pivot scale agree with the observational constraints \re{A}--\re{r}, except that $n_s$ can have the somewhat wider range $[0.95,0.98]$, and the number of e-folds until the end of inflation agrees with \re{Nval} to within $\pm1$. We restrict the Higgs quartic coupling to the range $[10^{-5},10^{-1}]$. Due to loop corrections, $\lambda$ runs to smaller values with increasing scale and can even cross zero. Therefore, it can be arbitrarily small at the pivot scale, but very small values require tuning, and the running can spoil the flatness of the potential. (This happens in the minimally coupled case \cite{Isidori:2007vm, Hamada:2013mya, Fairbairn:2014nxa}.)

\begin{figure}[t!]
\centering
\includegraphics[scale=0.55]{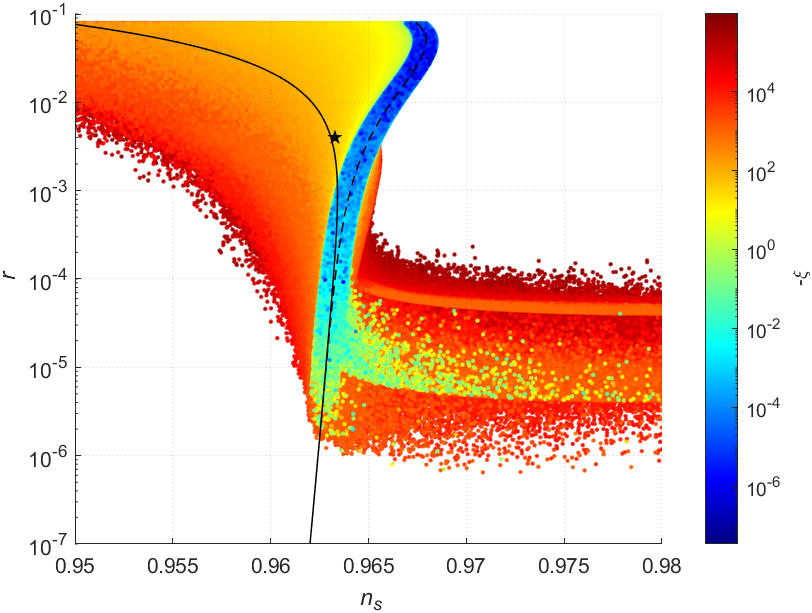}
\includegraphics[scale=0.55]{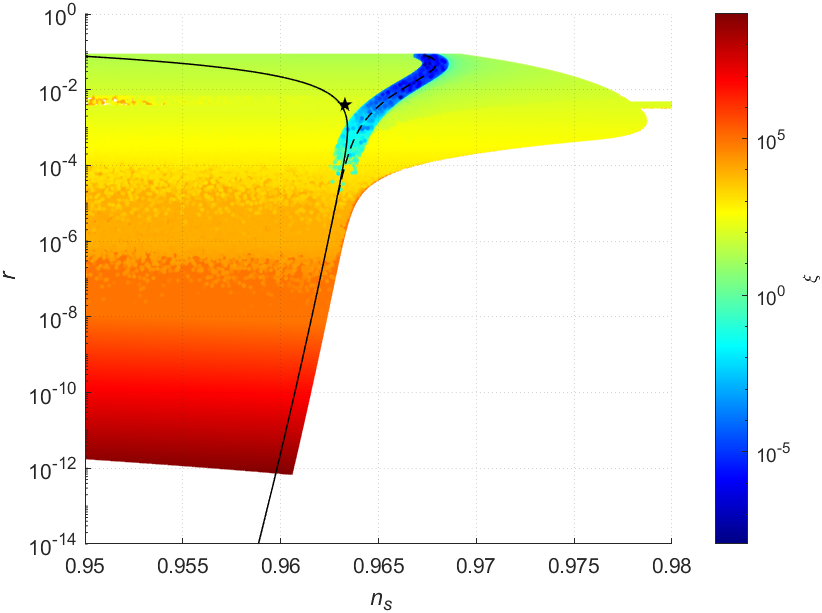}
\caption{Spectral index $n_s$ and tensor-to-scalar ratio $r$ for $\xi<0$ (left) and $\xi>0$ (right). The colour corresponds to the smallest absolute value of the non-minimal coupling $|\xi|$ to the Ricci scalar. The points satisfy all observational CMB constraints, except that the range of $n_s$ is wider. The solid line traces the prediction of plateau inflation, and the dashed line is the case $\xi=0$. The star marks the metric case.}
\label{fig:nsr}
\end{figure}

In \fig{fig:nsr} (left) we show the results on the $(n_s,r)$ plane. The colour indicates the smallest value of $|\xi|$ (it is not single-valued on this plane). The solid line is the analytical result \re{plateau} for plateau inflation, and the dashed line is the result \re{xi0results} for $\xi=0$. (It lies in the middle of the blue region corresponding to the limit $\xi\to0$ because in the numerical scan we allow a variation $\pm1$ in $N$.) The star marks the metric case.

The tensor-to-scalar ratio extends from the maximum observationally allowed value down to around $10^{-6}$, and $n_s$ covers the entire current observational range. The running is in the small range $-1\times10^{-3}\lesssim\a\lesssim-5\times10^{-4}$. The running of the running also has a small range, $-6\times10^{-5}\lesssim\b\lesssim-2\times10^{-5}$. The non-minimal couplings have the ranges $|\xi|<10^6$, $|H_0|<10^8$ and $|Y_1|>3\times10^3$; $H_1$ can take any value in the range we scan.

Smooth, well-defined edges in the figures correspond to true observational constraints for the cosmological observables, while rough edges with individual points scattered about correspond to regions of parameter space that the scan has not fully resolved. In such regions the parameter values for points that satisfy the observational constraints are highly tuned, requiring the precise cancellation of two or more large numbers.

\subsection{$\xi>0$} \label{sec:xipos}

Finally, let us discuss the case when we include all three coupling functions and $\xi>0$. As in the case $\xi<0$, inflection point inflation is possible. Successful inflation is now possible also when $H_0=0$. We perform the same kind of numerical scan as in the case $\xi<0$, except the range of $\xi$ is now $[0, 10^{10}]$. In \fig{fig:nsr} we show the results on the $(n_s,r)$ plane.

The range of the predictions extends to much lower values of $r$ than in the case $\xi<0$. All of the edges of the allowed region are now well resolved. The non-minimal coupling of the Ricci scalar takes values $\xi<1\times10^9$; $H_0$, $H_1$ and $Y_1$ can take any value in the range we scan. In contrast to the case when one of the three couplings $\xi$, $H$ and $Y$ is zero, the predictions for $n_s$ and $r$ cover almost all of the range expected to be tested by next generation CMB experiments such the Simons Observatory \cite{Abitbol:2019nhf}, LiteBIRD \cite{Sugai:2020pjw} and CMB-S4 \cite{Abazajian:2019tiv}. However, there are regions on the $(n_s,\a)$ and $(n_s,\beta)$ planes with both positive and negative running within reach of upcoming experiments that the model cannot reproduce.

\section{Conclusions} \label{sec:conc}

\para{New perspective on inflation in LQG.}

We have studied the effect of non-minimal coupling of a scalar field to the Holst and Nieh--Yan terms on inflation, in addition to the non-minimal coupling to the Ricci scalar. These terms play a key role in LQG, and are expected to appear in theories where torsion is non-zero. Since the Higgs exists, it will in general couple to these terms, and the couplings have to be taken into account. Motivated by Higgs inflation, we have included terms up to dimension 4 and even in the field.

Non-minimal coupling to the Holst term alone gives inflation with predictions close to those of the metric formulation plateau Higgs inflation for the same amount of e-folds, although reheating and hence the number of e-folds may be different due to the different shape of the effective potential \cite{Bezrukov:2008ut, GarciaBellido:2008ab, Figueroa:2009, Figueroa:2015, Repond:2016, Ema:2016, DeCross:2016, Sfakianakis:2018lzf, Rubio:2019}. This means that observational verification of the predictions of this simplest metric formulation Higgs inflation \cite{Bezrukov:2007} would not rule out the Palatini formulation of Higgs inflation. That prediction has been earlier reproduced in the Palatini case with tuned non-metricity terms \cite{Rasanen:2018b}, but the present case shows it can be achieved with a simple Higgs-LQG action with no tuning. Adding a non-minimal coupling $\xi$ to the Ricci scalar recovers the results of Higgs plateau inflation in the Palatini formulation \cite{Bauer:2008} unless the Holst coupling $H_0$ is much larger than $|\xi|$. If the Holst term coupling dominates but $|\xi|$ also contributes, the spectral index $n_s$ and its running can be adjusted from the metric case. Notably, this form of inflation is possible even when $\xi$ is negative.

A non-minimal coupling to the Nieh--Yan term alone does not give successful inflation. If we also have $\xi\neq0$, plateau inflation can be modified so that it interpolates between the results we get in the Palatini and the metric formulation when only $\xi$ is non-zero, and the tensor-to-scalar ratio $r$ can be even larger than in the metric case. This case is identical to Higgs inflation in the teleparallel formulation \cite{Raatikainen:2019}.

If we include non-minimal coupling to all three terms (Ricci scalar, Holst term and Nieh--Yan term), the range of predictions for $n_s$ and $r$ widens considerably to cover almost all of the values expected to be covered by near-future experiments. However, when we add running or running of the running, not all values to be probed can be reproduced. Also, many of the values correspond to tuned couplings. For example, we can a produce an inflection point, but this requires carefully adjusting the non-minimal couplings, as has been done with quantum corrections \cite{Allison:2013uaa, Bezrukov:2014bra, Hamada:2014iga, Bezrukov:2014ipa, Rubio:2015zia, Fumagalli:2016lls, Enckell:2016xse, Bezrukov:2017dyv, Rasanen:2017, Masina:2018ejw, Salvio:2017oyf, Ezquiaga:2017fvi, Rasanen:2018a} and classical non-metricity terms \cite{Rasanen:2018b}.

It is interesting that the Higgs field makes the Holst and Nieh--Yan terms dynamical at the classical level, as fermions have been found to do \cite{Freidel:2005sn, Randono:2005, Perez:2005, Mercuri:2006um, Mercuri:2006wb, Bojowald:2007nu, Kazmierczak:2008}. The Higgs generates torsion, which makes the Holst term non-zero. The Holst term can have a large impact on inflation even if it is minimally coupled as long as either the Ricci scalar or the Nieh--Yan term have non-minimal coupling. However, the value for the minimal Holst term coupling (\ie the Barbero---Immirzi parameter) $1/\c\approx3.6$ determined from black hole entropy in the case with no chemical potential \cite{Ghosh:2004}, is too small to be discernible from the CMB.

The non-minimal couplings to the Higgs provide a new point of view on LQG cosmology. Just as a large $\xi$ brings the gravity scale down, so that (in the Jordan frame) gravitons violate perturbative unitarity below the Planck scale \cite{Barbon:2009, Burgess:2009, Burgess:2010zq, Lerner:2009na, Lerner:2010mq, Hertzberg:2010, Bauer:2010, Bezrukov:2010, Bezrukov:2011a, Calmet:2013, Weenink:2010, Lerner:2011it, Prokopec:2012, Xianyu:2013, Prokopec:2014, Ren:2014, Escriva:2016cwl, Fumagalli:2017cdo, Gorbunov:2018llf, Ema:2019, Bauer:2010, McDonald:2020, Shaposhnikov:2020fdv, Enckell:2020lvn}, large values of the non-minimal couplings of the Holst and Nieh--Yan terms can bring aspects of LQG down to the scales probed during inflation. They could also help address the  issue of apparent violation of unitarity, whose scale is known to be sensitive to the form of the kinetic term \cite{Bauer:2010, McDonald:2020, Shaposhnikov:2020fdv, Enckell:2020lvn}, which is affected by the Holst and Nieh--Yan terms.

\acknowledgments

ML would like to thank E. Wilson-Ewing for worthwhile suggestions. We thank Eemeli Tomberg for help with renormalisation group equations.

\appendix

\section{Solving for torsion in the tetrad formalism} \label{sec:tet}

As the tetrad formalism is more familiar to the LQG community, we cover briefly how the results \eqref{Gammasol}--\eqref{qt} for torsion can be elegantly obtained using tetrads. A set of tetrads $\{e^A{}_\alpha\}$ is a linear map from tangent space to spacetime, providing a basis for the tangent space at each point in spacetime; capital Latin indices are associated to the tangent space. We take the basis to be orthonormal with respect to the metric $g_{\alpha\beta}$,
\bea
  g_{\alpha\beta} = \eta_{A B} e^A{}_\alpha e^B{}_\beta \ ,
\eea
where $\eta_{A B}=\diag(-1,1,1,1)$ is the Minkowski metric. We also have $g^{\a\b} e^A{}_\alpha e^B{}_\beta = \eta^{AB}$. The inverse tetrad $e_A{}^\alpha$ is defined so that $e_A{}^\alpha e^A{}_\beta = \delta^\alpha{}_\beta$ and $e_A{}^\alpha e^B{}_\alpha = \delta^B{}_A$.
We assume from the beginning that the full covariant derivative of the tetrad, acting on spacetime and tangent space indices, vanishes \ie that the tetrad postulate holds.

In terms of tetrads and the tangent space connection $\omega_\a{}^{AB}$ (called the Lorentz connection), the action \eqref{actionJ} reads
\begin{equation}\label{eq:componentholstform}
\begin{split}
S =& \int \rmd^{4}x e\Bigg[ \ha F(h) e_A{}^\a e_B{}^\b F_{\a\b}{}^{AB} + \frac{1}{4} H(h) \epsilon\indices{_{AB}^{CD}} e_C{}^\a{} e_D{}^\b F_{\a\b}{}^{AB} - \frac{1}{4} Y(h) \epsilon^{\a\b\c\d}\eta_{AB}T^{A}{}_{\a\b}T^{B}{}_{\c\d}\\& - \ha K(h) \eta^{AB} e_A{}^\a e_B{}^\b \partial_{\a}h\, \partial_{\b} h - V(h)\Bigg] \ ,
\end{split}
\end{equation}
where $e=\det{(e^A{}_\alpha)}$, $F_{\a\b}{}^{AB}=2\pat_{[\a}\omega_{\b]}{}^{AB}+2\omega_{[\a}{}^{AC} \omega_{\b]}{}^{DB} \eta_{CD}$ is the curvature of the Lorentz connection, related to the Riemann tensor via $e^{\mu}_{\,\,\,A} e^{\nu}_{\,\,\, B} F^{\,\,\,\,\,\,AB}_{\a\b} = R^{\mu\nu}_{\,\,\,\,\,\, \a\b}$. Torsion is defined as $T^{A}{}_{\a\b}=\mathcal{D}_{[\a}e^{A}{}_{\b]}$, where the covariant derivative $\mathcal{D}$ acts only on tangent space indices.

The Einstein--Hilbert term plus the Holst term, together known as the Holst action, can be written compactly as
\begin{equation}\label{eq:componentholstform1}
S_{\textrm{Holst}}=\int \rmd^{4}x e\, \ha e_A{}^\a e_B{}^\b P\indices{^{AB}_{CD}} F_{\a\b}{}^{CD} \ ,
\end{equation}
where the projection operator is defined as
\begin{equation}\label{P}
P\indices{^{AB}_{CD}}= F\delta^{[A}{}_{C}\delta^{B]}{}_{D}+\frac{1}{2}H\epsilon\indices{^{AB}_{CD}} \ .
\end{equation}
The inverse of the projection operator is
\begin{equation}\label{eq:invoperatorphi}
(P^{-1})\indices{^{AB}_{CD}}=\frac{1}{F^2+H^2}\left( F\delta^{[A}{}_{C}\delta^{B]}{}_{D}-\frac{1}{2}H\,\epsilon\indices{^{AB}_{CD}}\right).
\end{equation}
Varying the action \eqref{eq:componentholstform} with respect to the Lorentz connection and dropping a boundary term gives the equation of motion
\begin{equation} \label{eomtetrad1}
\frac{1}{4}\mathcal{D}_{\a}\left(P\indices{^{AB}_{CD}}\epsilon_{ABEF}\epsilon^{\a\b\c\d}e^{E}{}_{\c} e^{F}{}_{\d}\right)+\frac{1}{2} \epsilon_{CDEF}e^{E[\a}e^{\vert F\vert \b]} \partial_{\a} Y = 0 \ .
\end{equation}
The solution to \re{eomtetrad1} is obtained straightforwardly, using the definition of torsion and operating with \eqref{eq:invoperatorphi}: 
\begin{equation}\label{eq:tetradtorsion7}
T\indices{^A_{\a\b}}=\frac{1}{F^2+H^{2}}\bigg\{ e^A{}_{[\a} \big[(F\pat_{\b]}F+H(\pat_{\b]}H-\pat_{\b]}Y)\big] + e^A{}_{\c} \epsilon\indices{^{\c}_{\a\b}^\d} \big[H\pat_{\d} F-F(\pat_{\d} H-\pat_{\d}Y) \big] \bigg\} \ .
\end{equation}
This agrees with the solution for torsion in \eqref{Gammasol}--\eqref{qt}.

\bibliographystyle{JHEP}
\bibliography{ash}

\end{document}